\documentclass{fmcad}

\usepackage{cite}

\usepackage{soul}              
\usepackage{url}               
\usepackage[utf8]{inputenc}   

\usepackage{amsmath,amsfonts,amssymb}  
\usepackage{xspace}

\usepackage{macro}

\def\1{\bm{1}}

\DeclareMathAlphabet{\mathsfit}{\encodingdefault}{\sfdefault}{m}{sl}
\SetMathAlphabet{\mathsfit}{bold}{\encodingdefault}{\sfdefault}{bx}{n}

\newcommand{\defeq}{\triangleq}

\newcommand{\funcdep}{\rightarrow}

\newcommand{\vmin}{v_{\text{min}}}
\newcommand{\vmax}{v_{\text{max}}}

\usepackage{graphicx}          
\usepackage{xurl}                
\usepackage{booktabs}            
\usepackage{nicefrac}            
\usepackage{microtype}           
\usepackage{xcolor}              
\usepackage{comment}           
\usepackage{listings}            
\usepackage{wrapfig}             
\usepackage{enumitem}           
\usepackage{algorithm}           
\usepackage{algorithmic}        
\usepackage{cleveref}  

\lstdefinelanguage{SQL}{
  keywords={SELECT, FROM, WHERE, JOIN, ON, AND, OR, NOT, NULL, 
            DISTINCT, COUNT, MIN, MAX, GROUP, BY, ORDER, ASC, DESC,
            BETWEEN, IN, AS, LIMIT},
  keywordstyle=\color{blue}\bfseries,
  sensitive=false,
  comment=[l]{--},
  commentstyle=\color{gray}\ttfamily,
  stringstyle=\color{red}\ttfamily,
  morestring=[b]',
  morestring=[b]"
}

\usepackage{peanutgallery}

\title{\spotitplus: Verification-based Text-to-SQL Evaluation with Database Constraints}
\author{
    \IEEEauthorblockN{
        Andrew Tremante\IEEEauthorrefmark{1},
        Rocky Klopfenstein\IEEEauthorrefmark{1},
        Yang He\IEEEauthorrefmark{2},
        Yuepeng Wang\IEEEauthorrefmark{2},
        Nina Narodytska\IEEEauthorrefmark{3},
        Haoze Wu\IEEEauthorrefmark{1}\IEEEauthorrefmark{3}
    }

    \IEEEauthorblockA{
        \IEEEauthorrefmark{1}Amherst College,
        \IEEEauthorrefmark{2}Simon Fraser University,
        \IEEEauthorrefmark{3}VMware Research by Broadcom
    }
}

\begin{document}

\maketitle

\begin{abstract}
We present \spotitplus, an open-source tool for evaluating \txttosql systems via bounded equivalence verification. Given a generated SQL query and the ground truth, \spotitplus actively searches for database instances that differentiate the two queries. To ensure that the generated counterexamples reflect practically relevant discrepancies, we introduce a best-effort constraint-mining pipeline that combines rule-based specification mining with LLM-based validation over example databases.
Experimental results on the BIRD dataset show that the mined constraints enable \spotitplus to generate more realistic differentiating databases, while preserving its ability to efficiently uncover numerous discrepancies between generated and gold SQL queries that are missed by standard test-based evaluation.
\end{abstract}

\section{Introduction}

Text-to-SQL\,--\,the task of translating natural language questions into executable database queries\,--\,constitutes a key building block of modern chatbots and intelligent assistants across a wide range of industrial applications~\cite{amazon,msd2024,bitsai,splunk}. To compare and track the capabilities of current \txttosql methods, several community-driven evaluation platforms have been proposed~\cite{lei2024spider,birdleader}. These platforms are widely endorsed by both industry and academia as primary benchmarks for assessing the state of the art in \txttosql.

Determining whether a \txttosql method produces a correct SQL query is non-trivial. Existing platforms predominantly rely on \emph{test-based evaluation}: executing the generated SQL query and the human-labeled ground truth (i.e., the gold SQL) on a fixed test database and comparing the results. This approach can be overly optimistic, as two non-equivalent queries may return identical results on test database instances. Recent work~\cite{spotit} demonstrates that test-based evaluation does frequently overlook such discrepancies and proposes a verification-based alternative that provides stronger correctness guarantees. The key idea is to use an SMT-based bounded verification engine~\cite{verieql} to systematically search for database instances that differentiate the generated and gold queries. The outcome is either a proof of equivalence within the bounded search space, or a concrete counterexample database witnessing non-equivalence.

A natural concern with verification-based evaluation is the practical relevance of the discovered counterexamples. In addition to explicitly stated integrity constraints (e.g., primary keys and foreign key relationships), databases often contain domain-specific constraints that are implicit yet should be obeyed from a practical standpoint. With these constraints unexposed to the underlying verifier, the found counterexample databases might reveal discrepancies that only occur in pathological corner cases that are impossible or unlikely to arise in practice. 
Although the ultimate assessment of realism rests with human judgment, that is, the user is ultimately responsible and uniquely eligible to decide which constraints best reflect their domain, we argue that a practical \txttosql evaluation framework should strive to ensure that counterexamples reflect plausible data distributions whenever possible.

In this paper, we present \spotitplus, a bounded-verification-based tool for \txttosql evaluation. \spotitplus goes beyond an open-source implementation of the verification-based pipeline described in prior work~\cite{spotit}. It features a constraint-extraction pipeline that mines database constraints from example databases. These constraints are encoded as additional SMT constraints during bounded verification, effectively augmenting the integrity constraints in the underlying verification formula and restricting the solver's search to database instances that respect realistic data distributions. 

Purely rule-based constraint mining, however, risks overfitting to idiosyncrasies of the example databases. 
To mitigate this risk, \spotitplus incorporates a large language model (LLM) to assess whether a mined constraint represents a genuine domain property before including it in the verification query. Note that constraint extraction is inherently a best-effort procedure, and \spotitplus does not claim to fully solve the challenge of realistic \txttosql evaluation; rather, our goal is to provide configurable and practical additions to existing \txttosql evaluation options.

We employ \spotitplus to evaluate ten state-of-the-art \txttosql methods on the popular BIRD dataset~\cite{birdleader}. Our results show that incorporating the extracted database constraints yields more realistic counterexamples, while still uncovering a substantial number of discrepancies that test-based evaluation overlooks. To summarize, our contributions include:
\begin{itemize}[topsep=3pt, itemsep=2pt, parsep=0pt]
    \item \spotitplus, an open-source verification-based tool\footnote{Available at \url{https://github.com/ai-ar-research/SpotIt-plus}} for \txttosql evaluation;
    \item A constraint-extraction pipeline that combines rule-based specification mining with LLM-based validation and repair;
    \item An empirical evaluation demonstrating that \spotitplus improves the realism of counterexamples while preserving strong discrepancy-detection capability.
\end{itemize}

\section{Preliminaries}
\label{app:background}

Determining whether two SQL queries produce identical results on all possible database instances is generally undecidable for expressive SQL fragments~\cite{chu2018axiomatic,chu2017hottsql,wang2024qed,zhou2024solving,zhou2022spes}. Bounded equivalence verification addresses this by checking equivalence within a finite search space: given queries $Q_1$ and $Q_2$ over schema $S$ and bound $K$, the goal is to determine whether $Q_1$ and $Q_2$ are equivalent on all databases where each relation has at most $K$ tuples~\cite{chu2017demonstration,chu2017cosette,verieql,veanes2010qex}. Formally:

\begin{align*}
Q_1 \simeq_{S,K} Q_2 \defeq\ &\forall D \in \text{Instances}(S).\ \forall R \in \text{Relations}(D). \\
&|R| \leq K \implies Q_1(D) = Q_2(D)
\end{align*}

If $Q_1 \simeq_{S,K} Q_2$ holds, the queries are deemed \emph{boundedly equivalent}; otherwise, a bounded equivalence checker produces a concrete counterexample database $D_{\text{cex}}$ demonstrating $Q_1(D_{\text{cex}}) \neq Q_2(D_{\text{cex}})$.

\subsection{SMT-based Bounded Equivalence Checking}

Bounded SQL equivalence verification can be reduced to satisfiability modulo theories (SMT). Both queries are symbolically executed on a database with $K$ symbolic tuples per relation, and the execution together with a non-equivalence assertion is encoded as an SMT formula:
\[
\Phi = \Phi_{C} \land \Phi_{Q_1} \land \Phi_{Q_2} \land \neg\Phi_{\Leftrightarrow}
\]
where $\Phi_{C}$ encodes database integrity constraints (e.g., primary and foreign keys), $\Phi_{Q_1}$ and $\Phi_{Q_2}$ encode the semantics of each query, and $\neg\Phi_{\Leftrightarrow}$ asserts that the two query outputs differ. If $\Phi$ is unsatisfiable, the queries are boundedly equivalent; otherwise, a satisfying assignment yields a counterexample database.

For example, consider verifying \texttt{SELECT id FROM R WHERE id > 1} against \texttt{SELECT id FROM R WHERE id > 2} with $K=1$. \verieql creates a symbolic tuple $t_1 = [x_1, x_2]$ for relation $R$ and symbolic result tuples $Q_1(D) = [t_2]$ where $t_2 = [x_3]$ and $Q_2(D) = [t_3]$ where $t_3 = [x_4]$. Query semantics are captured as:
\[
\Phi_{Q_1} = (x_1 > 1 \to (x_3 = x_1 \land \neg\text{Del}(t_2))) \land (x_1 \leq 1 \to \text{Del}(t_2))
\]
where $\text{Del}$ is a deletion flag indicating tuple non-existence. The conjunction $\Phi_{Q_1} \land \Phi_{Q_2} \land (t_2 \neq t_3)$ is satisfiable if and only if a differentiating database exists.

For the equivalence predicate $\Phi_{\Leftrightarrow}$, SQL equivalence checkers may adopt bag, list, or set semantics. \spotitplus uses \emph{set equivalence}, in accordance with the convention in existing \txttosql evaluations~\cite{birdleader,lei2024spider}; this choice can be adjusted to match other evaluation settings.

Prior work encodes in $\Phi_C$ only the integrity constraints explicitly stated in the database schema. The key contribution of \spotitplus is to \emph{augment} $\Phi_C$ with additional domain constraints automatically mined from example databases, thereby restricting the solver's search to realistic database instances in a best-effort manner and producing counterexamples that reflect plausible data distributions.

In practice, \spotitplus uses \verieql~\cite{verieql} as its underlying bounded equivalence checker. \verieql encodes SQL semantics symbolically via \zthree, and supports a rich SQL subset including joins, aggregations, subqueries, and set operations with user-defined integrity constraints such as primary keys, foreign keys, and uniqueness constraints. To the best of our knowledge, it supports the most expressive SQL fragment among existing bounded equivalence checkers.

\section{Motivating Example}
\label{sec:motivation}

Before presenting the workflow of \spotitplus, we first illustrate the effect of constraint extraction and LLM validation through a motivating example. Consider the following question from the BIRD dataset~\cite{birdleader}: 
\emph{Please list the account types that are not eligible for loans, and the average income of residents in the district where the account is located exceeds \$8000 but is no more than \$9000.} Consider the generated query and the gold query for this question in Figure~\ref{fig:example-sql}.

\begin{figure}[t!]
\centering
\begin{lstlisting}[language=SQL, basicstyle=\footnotesize\ttfamily, breaklines=true, numbers=none]
/* Generated Query */
SELECT DISTINCT DISP.TYPE 
FROM DISP 
INNER JOIN ACCOUNT 
  ON DISP.ACCOUNT_ID = ACCOUNT.ACCOUNT_ID 
INNER JOIN DISTRICT 
  ON ACCOUNT.DISTRICT_ID = DISTRICT.DISTRICT_ID 
WHERE DISP.TYPE <> 'OWNER' 
  AND DISTRICT.A11 > 8000 
  AND DISTRICT.A11 <= 9000;
\end{lstlisting}
\begin{lstlisting}[language=SQL, basicstyle=\footnotesize\ttfamily, breaklines=true, numbers=none]
/* Gold Query */
SELECT T3.TYPE 
FROM DISTRICT AS T1 
INNER JOIN ACCOUNT AS T2 
  ON T1.DISTRICT_ID = T2.DISTRICT_ID 
INNER JOIN DISP AS T3 
  ON T2.ACCOUNT_ID = T3.ACCOUNT_ID 
WHERE T3.TYPE != 'OWNER' 
  AND T1.A11 BETWEEN 8000 AND 9000;
\end{lstlisting}
\vspace{-8pt}
\caption{The generated and gold queries for a question in the BIRD dataset.}
\label{fig:example-sql}
\vspace{-10pt}
\end{figure}

\begin{figure*}[t!]
    \centering
    \includegraphics[width=0.75\textwidth]{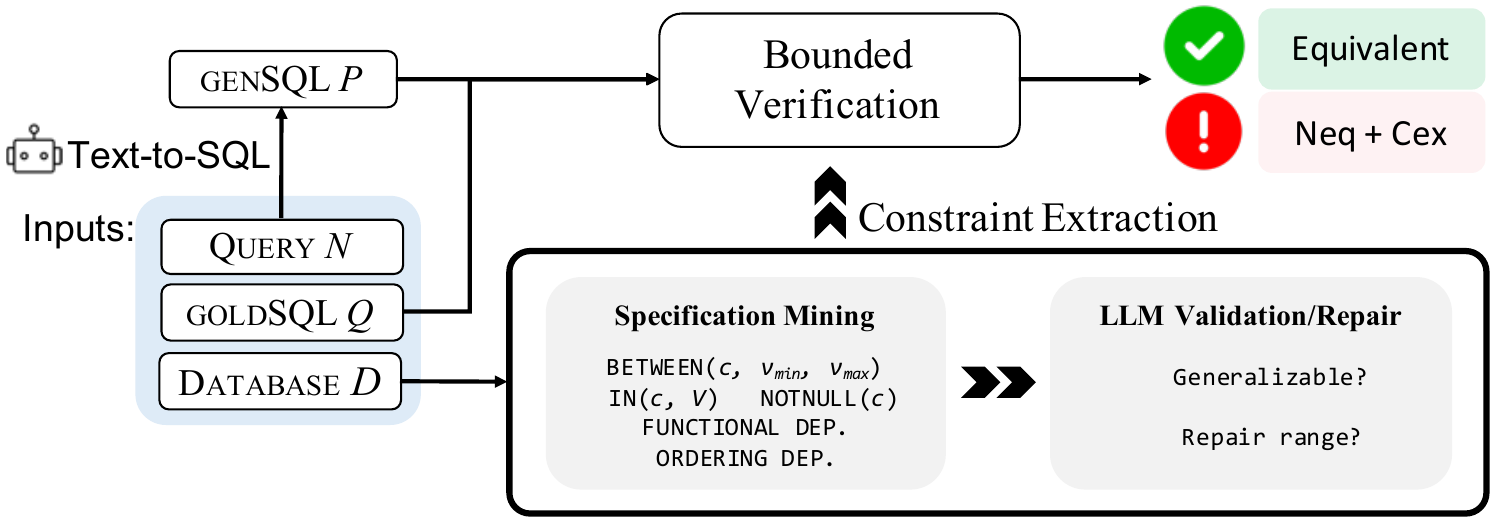}
    \caption{The workflow of \spotitplus. The `Equivalent' outcome denotes bounded equivalence up to $K$ rows per table.}
    \label{fig:extraction-pipeline}
\end{figure*}

These queries differ in how they filter the column \texttt{DISTRICT.A11}. The generated query includes only values strictly greater than 8000 (\texttt{DISTRICT.A11 > 8000}), while the gold query uses \texttt{BETWEEN 8000 AND 9000}, which includes 8000.\footnote{In this case, the gold SQL is incorrect, which is a separate issue discussed in prior work~\cite{spotit}.} 
The official test-based evaluation incorrectly deems these two queries equivalent because they produce the same result on the test database. In contrast, bounded verification without extracted database constraints finds the following counterexample that distinguishes the two queries:
\begin{lstlisting}[language=SQL, basicstyle=\scriptsize\ttfamily, breaklines=true, numbers=none]
--SpotIt Counterexample--
CREATE TABLE DISTRICT (DISTRICT_ID INTEGER, A2 VARCHAR(20), A3 VARCHAR(20), ..., A11 INTEGER, ...);
INSERT INTO DISTRICT VALUES (0, '2147483648', '2147483648', ..., 8000, ...);
CREATE TABLE ACCOUNT (ACCOUNT_ID INTEGER, DISTRICT_ID INTEGER, FREQUENCY VARCHAR(20), ...);
INSERT INTO ACCOUNT VALUES (0, 0, '2147483648', ...);
INSERT INTO ACCOUNT VALUES (1, 0, '2147483648', ...);
CREATE TABLE DISP (DISP_ID INTEGER, CLIENT_ID INTEGER, ACCOUNT_ID INTEGER, TYPE VARCHAR(20));
INSERT INTO DISP VALUES (0, 0, 0, '2147483648');
INSERT INTO DISP VALUES (1, 0, 1, '556585159121467050');
\end{lstlisting}
The counterexample generated by \spotitplus with analytically extracted and LLM-validated database constraints is as follows:
\begin{lstlisting}[language=SQL, basicstyle=\scriptsize\ttfamily, breaklines=true, numbers=none]
--SpotIt+ Counterexample--
CREATE TABLE DISTRICT (DISTRICT_ID INTEGER, A2 VARCHAR(20), A3 VARCHAR(20), ..., A11 INTEGER, ...);
INSERT INTO DISTRICT VALUES (1, '2147483648', 'Prague', ..., 8000, ...);
CREATE TABLE ACCOUNT (ACCOUNT_ID INTEGER, DISTRICT_ID INTEGER, FREQUENCY VARCHAR(20), ...);
INSERT INTO ACCOUNT VALUES (1, 1, 'POPLATEK TYDNE', ...);
INSERT INTO ACCOUNT VALUES (2, 1, 'POPLATEK PO OBRATU', ...);
CREATE TABLE DISP (DISP_ID INTEGER, CLIENT_ID INTEGER, ACCOUNT_ID INTEGER, TYPE VARCHAR(20));
INSERT INTO DISP VALUES (1, 1, 1, 'DISPONENT');
INSERT INTO DISP VALUES (2, 1, 1, 'DISPONENT');
\end{lstlisting}

The counterexample generated by \spotitplus is qualitatively more realistic. In particular, categorical constraints were extracted for columns such as \texttt{DISTRICT.A2}, \texttt{ACCOUNT.FREQUENCY}, and \texttt{DISP.TYPE}, which state that the column values must belong to one of several fixed choices.

On the other hand, if we add all constraints mined directly from the test database without LLM-based validation, bounded verification fails to find a counterexample in this case. This is because, in the test database, the value of \texttt{DISTRICT.A11} ranges from 8110 to 12541 and does not include the critical boundary value 8000. Adding this range constraint would eliminate any potential counterexample. Indeed, this constraint is overly restrictive, and the LLM validation pass relaxes it.

\section{\spotitplus: Bounded Equivalence Verification with Database Constraints}
\label{sec:approach}

The workflow of \spotitplus is shown in Fig.~\ref{fig:extraction-pipeline}. A \txttosql benchmark consists of a natural language question $\queryshortnl$, the gold SQL $\goldqueryshort$, and an example database $\db$. A \txttosql method takes as input $\queryshortnl$ and generates a SQL query \genqueryshort. The goal of \spotitplus is to determine whether \genqueryshort should be deemed a correct prediction, and it does so by checking its bounded equivalence with the gold SQL \goldqueryshort under database constraints. 
Unlike previous work~\cite{spotit} that only considers integrity constraints explicitly stated in the database schema, \spotitplus features a constraint extraction module which takes as input the example database \db, mines constraints from it, and validates/repairs the constraints using an LLM. Those constraints are encoded in the bounded verification query to rule out unrealistic counterexamples. Background on bounded SQL equivalence checking can be found in App.~\ref{app:background}. Here, we focus on the constraint extraction module.

\subsection{Constraint Types}
\label{sec:constraint-types}

Currently, \textsc{SpotIt+} extracts five types of constraints from the example databases.

\bfpara{\range Constraints.}
\range constraints restrict a numeric column to values within a specified interval:
\[
\cbetween(c, \vmin, \vmax) \defeq \forall i.\ \vmin \leq c[i] \leq \vmax
\]
where $c$ is a column, $c[i]$ denotes the value of column $c$ at row $i$, and $\vmin$ and $\vmax$ define the lower and upper bounds.

\bfpara{\categorical Constraints.}
\categorical constraints restrict a column to a finite set of discrete values:
\[
\cin(c, V) \defeq \forall i.\ c[i] \in V
\]
where $V = \{v_1, v_2, \ldots, v_k\}$ is a finite set of allowed values.

\bfpara{\nnull Constraints.}
\nnull constraints specify that a column cannot contain null values:
\[
\cnotnull(c) \defeq \forall i.\ c[i] \neq \nullv
\]

\bfpara{\functional.}
\functional state that one column uniquely determines another column:
\[
c_1 \funcdep c_2 \defeq \forall i, j.\ c_1[i] = c_1[j] \implies c_2[i] = c_2[j]
\]
where $c_1$ and $c_2$ are individual columns, and $i,j$ range over all pairs of rows.

\bfpara{\ordering.}
\ordering enforce an inequality relationship between two numeric columns:
\[
c_1 \triangleleft c_2 \defeq \forall i.\ c_1[i] \triangleleft c_2[i]
\]
where $c_1$ and $c_2$ are numeric columns, $\triangleleft \in \{\leq, \geq\}$ is a comparison operator.

\begin{table*}[t!]
\centering
\small
\caption{Statistics of the BIRD development set~\cite{birdleader}.}
\label{tab:bird-stats}
\begin{tabular}{llcccc}
\toprule
\textbf{Database} & \textbf{Domain} & \textbf{Tables} & \textbf{Columns} & \textbf{Rows} & \textbf{Questions}\\
\midrule
\texttt{california\_schools} & Education & 3 & 89 & 9,980 & 89 \\
\texttt{card\_games} & Entertainment & 6 & 115 & 133,908 & 191 \\
\texttt{codebase\_community} & Technology & 8 & 71 & 92,581 & 186 \\
\texttt{debit\_card\_specializing} & Finance & 5 & 21 & 84,610 & 64 \\
\texttt{european\_football\_2} & Sports & 7 & 103 & 31,828 & 129 \\
\texttt{financial} & Finance & 8 & 55 & 134,960 & 106 \\
\texttt{formula\_1} & Sports & 13 & 95 & 39,561 & 174 \\
\texttt{student\_club} & Education & 8 & 48 & 5,314 & 158 \\
\texttt{superhero} & Entertainment & 10 & 31 & 1,061 & 129 \\
\texttt{thrombosis\_prediction} & Healthcare & 3 & 64 & 5,317 & 163 \\
\texttt{toxicology} & Science & 4 & 11 & 12,453 & 145 \\
\bottomrule
\end{tabular}
\end{table*}

\subsection{Constraint Extraction Pipeline}
\label{sec:extraction}

\spotitplus takes as input an example database (which in reality is provided as part of the official benchmark set) and automatically extracts the five types of constraints on all eligible columns in the database.
\range constraints are computed for each column of numerical type by computing the min and max values in the database. \categorical constraints are extracted for all columns with at most $K$ unique values, where each unique value is considered to be a category. $K$ is a configurable parameter and is by default set to 30. 
\nnull constraints are generated for columns without nulls in the test database.
\functional are extracted for all non-primary column pairs $c_1, c_2$ that satisfy the functional dependency relation. \ordering are extracted for non-primary, numerical column pairs $c_1, c_2$ that satisfy an inequality relation.

Next, \spotitplus uses an LLM to filter the inferred candidate constraints and repair over-constraining range constraints. Specifically, for each extracted constraint, a prompt that asks the LLM to evaluate whether the constraint holds beyond the test database is formulated. For example, a functional dependency between \texttt{country code} and \texttt{country name} represents a genuine relationship, while a dependency between \texttt{firstname} and \texttt{occupation} does not. For numerical columns, the LLM is further prompted to judge whether the extracted range constraint is too restrictive and should be repaired. For example, suppose the test database only contains \texttt{patient age} ranging from 30 to 60. The LLM may suggest to change the range constraint to $[0, 120]$.

The extracted database constraints are then encoded in the bounded verification engine to focus the search on practically relevant data distribution. App.~\ref{app:spotitplus} includes additional implementation details on the constraint extraction module and provides a sample system prompt for LLM validation.

Note that it is ultimately up to the user (e.g., Text-to-SQL developers who need to evaluate their methods, Text-to-SQL users who needs to choose between methods) to decide which constraints should be included; our goal is to provide configurable levels of strictness for equivalence checking and allow users to choose the most appropriate for their needs.

\section{Evaluation}
\label{sec:evaluation}

We use \spotitplus to evaluate \txttosql methods on the popular \bird dataset~\cite{birdleader}, whose publicly available dev-set contains 1,534 questions across 11 databases spanning healthcare, education, and other professional domains. We obtained predictions from 10 state-of-the-art \txttosql methods (shown in Table~\ref{tab:sqlmethods}) on the BIRD leaderboard.

\begin{table}[h!]
\footnotesize
\caption{The considered \txttosql methods.}
\centering
\label{tab:sqlmethods}
\begin{tabular}{cc}
\toprule
Entry & Acronym\\
\midrule
Alpha-SQL + Qwen2.5-Coder-32B~\cite{li2025alpha} & \alphasql \\
CSC-SQL + Qwen2.5-Coder-7B~\cite{sheng2025csc} &\cscsqlsmall\\
CSC-SQL + XiYanSQL~\cite{sheng2025csc} & \cscsqlbig \\
GenaSQL-1~\cite{donder2025cheaper}  & \genaonesql \\
GenaSQL-2~\cite{donder2025cheaper} & \genatwosql \\
RSL-SQL + GPT-4o~\cite{cao2024rsl} & \rslsql \\
OmniSQL-32B\cite{li2025omnisql} & \omnimajsql \\
GSR (anonymous authors) & \gsrsql \\
$\text{CHESS}_{\text{IR+CG+UT}}$~\cite{talaei2024chess} & \chesssql \\
SLM-SQL + Qwen2.5-Coder-1.5B~\cite{sheng2025slm} & \slmsql \\
\bottomrule
\end{tabular}
 
\end{table}

We are primarily interested in the following questions:
\begin{itemize}
\item How do the accuracy change when we check SQL equivalence under additional database constraints?
\item Can \spotitplus efficiently evaluate prediction correctness with the additional database constraints?
\item Can the added constraints result in more realistic counterexample databases? 
\end{itemize}

\begin{table*}[t!]
\centering
\small
\caption{Performance of Text-to-SQL methods evaluated under \ex, \vanilla, \ruleBased, and \LLM on the 1533 BIRD-dev benchmarks.} 
\label{tab:res}
\begin{tabular}{p{1.5cm}cccccccc}  
\toprule
& \multicolumn{2}{c}{\textbf{\ex}} & \multicolumn{2}{c}{\textbf{\exVanilla}} &
\multicolumn{2}{c}{\textbf{\exRuleBased}} & \multicolumn{2}{c}{\textbf{\exLLM}} \\
\cmidrule(lr){2-3} \cmidrule(lr){4-5} \cmidrule(lr){6-7} \cmidrule(lr){8-9}
& Acc. (\%) & Rnk & Acc. (\%) & Rnk & Acc. (\%) & Rnk & Acc. (\%) & Rnk \\
\midrule
\cscsqlbig & 71.32 & 1 & 60.10 & 3 & 63.56 & 2 & 63.43 & 2 \\
\genatwosql & 70.53 & 2 & 60.89 & 1 & 63.75 & 1 & 63.56 & 1 \\
\alphasql & 69.36 & 3 & 56.91 & 6 & 61.21 & 5 & 60.56 & 5 \\
\genaonesql & 69.23 & 4 & 60.50 & 2 & 63.23 & 3 & 63.17 & 3 \\
\cscsqlsmall & 69.17 & 5 & 59.91 & 4 & 62.58 & 4 & 62.26 & 4 \\
\rslsql & 67.67 & 6 & 57.63 & 5 & 60.82 & 6 & 60.56 & 6 \\
\omnimajsql & 66.88 & 7 & 55.80 & 8 & 59.52 & 7 & 59.32 & 7 \\
\gsrsql & 66.49 & 8 & 55.93 & 7 & 59.13 & 8 & 58.87 & 8 \\  
\chesssql & 63.62 & 9 & 53.91 & 9 & 56.91 & 9 & 56.45 & 9 \\
\slmsql & 63.43 & 10 & 52.54 & 10 & 56.00 & 10 & 55.54 & 10 \\
\bottomrule
\end{tabular}
\end{table*}



We first evaluate the predictions of each method using BIRD’s official test-based execution accuracy metric (\ex), which compares the results of executing the generated and gold queries on a given test database. We then apply \spotitplus to predictions that are deemed correct by \ex. We consider three configurations: 1) \vanilla: bounded equivalence checking without extracted database constraints. This is the configuration used in prior work~\cite{spotit}; 2) \ruleBased: bounded equivalence checking with extracted database constraints, but without LLM validation and repair; and 3) \LLM: bounded verification with LLM-validated constraints.

Constraint extraction is performed offline once per database. Table~\ref{tab:bird-stats} summarizes the statistics of the 11 BIRD-dev databases. The specification mining pass took approximately 5 minutes to process all 11 databases sequentially. Validating the 6,264 mined constraints via the OpenAI API took approximately 2.5 hours, yielding 1,916 remaining constraints. These costs are reasonable as a one-time computation that can be further reduced through parallelization.

We verify each SQL pair up to a bound of 5 rows per table, prior work~\cite{spotit} suggests that the number of additional counterexamples is marginal for $K \geq 3$. Each pair is assigned one physical core, 8GB of memory, and a CPU timeout of 600 seconds. In practice, counterexamples are typically found within seconds, as discussed below. Experiments were conducted on a cluster of Dell PowerEdge R6525 servers equipped with 2.6GHz AMD CPU cores. An artifact that contains scripts to reproduce our experiments is publicly available~\cite{spotitplus-artifact}.

\subsection{Verification Results}

Table~\ref{tab:res} reports the accuracy of each Text-to-SQL method under the official test-based metric \ex and the three verification-based evaluation metrics. All verification-based configurations identify substantially more discrepancies between generated and gold SQL queries than the test-based evaluation.

Compared to \vanilla, both \ruleBased and \LLM deem more query pairs equivalent. This is expected as the extracted database constraints eliminate query discrepancies that arise only in unrealistic database instances. 
Interestingly, comparing \ruleBased and \LLM shows that LLM validation does not lead to a substantial change in the number of pairs deemed inequivalent. We hypothesize that most database constraints do not eliminate counterexamples, but only make them more realistic. Nonetheless, as shown in Sec.~\ref{sec:motivation}, LLM-based validation does recover genuine discrepancies from overly restrictive constraints sometimes. Appendix~\ref{app:examples} provides examples where \LLM rules out counterexamples found by \vanilla, and additional examples showing counterexamples generated with and without database constraints.

Table~\ref{tab:runtime} reports per-method runtime statistics for counterexample generation. Runtime performance remains efficient across all methods and configurations. The mean average time is 1.7 seconds for \vanilla, 1.4 seconds for \ruleBased, and 0.9 seconds for \LLM. Median runtimes stay below 0.6 seconds across all configurations, indicating that the majority of queries are verified within one second. The runtime reductions under \ruleBased and \LLM are attributable to the restricted search space imposed by the additional constraints. Currently, \verieql~\cite{verieql} successfully encodes 93--97\% of the examined SQL pairs across the ten \txttosql methods. While the coverage on BIRD benchmarks is already high, closing the remaining gap is an important next step.

\begin{table}[t!]
\centering
\caption{Average and median runtime (in seconds) for counterexample generation across Text-to-SQL methods.}
\label{tab:runtime}
\begin{tabular}{lcccccc}
\toprule
& \multicolumn{2}{c}{\textbf{\vanilla}} & \multicolumn{2}{c}{\textbf{\ruleBased}} & \multicolumn{2}{c}{\textbf{\LLM}} \\
\cmidrule(lr){2-3} \cmidrule(lr){4-5} \cmidrule(lr){6-7}
\textbf{Method} & Avg. & Med. & Avg. & Med. & Avg. & Med. \\
\midrule
\cscsqlbig  & 2.87 & 0.33 & 0.72 & 0.46 & 0.85 & 0.47 \\
\genatwosql & 1.06 & 0.30 & 2.26 & 0.36 & 0.98 & 0.34 \\
\alphasql   & 3.67 & 0.36 & 0.99 & 0.58 & 1.57 & 0.55 \\
\genaonesql & 0.82 & 0.39 & 0.66 & 0.40 & 0.57 & 0.37 \\
\cscsqlsmall& 0.58 & 0.29 & 3.08 & 0.36 & 1.37 & 0.34 \\
\rslsql     & 3.34 & 0.41 & 1.24 & 0.48 & 0.90 & 0.43 \\
\omnimajsql & 0.94 & 0.35 & 2.56 & 0.59 & 0.65 & 0.50 \\
\gsrsql     & 0.80 & 0.34 & 0.75 & 0.53 & 0.85 & 0.45 \\
\chesssql   & 1.26 & 0.28 & 0.75 & 0.41 & 0.67 & 0.39 \\
\slmsql     & 1.29 & 0.29 & 0.66 & 0.36 & 0.56 & 0.37 \\
\midrule
\textbf{Mean} & 1.66 & 0.33 & 1.37 & 0.45 & 0.90 & 0.42 \\
\bottomrule
\end{tabular}
\vspace{-0.3cm}
\end{table}

\section{Conclusion}
We presented \spotitplus, an \txttosql evaluation tool based on bounded equivalence verification and database constraint extraction. Currently, \spotitplus extracts  from example databases five types of constraints 
and uses an LLM to validate and repair the extracted constraints. Experimental evaluation demonstrated that \spotitplus is readily applicable to high-profile \txttosql evaluation platform such as BIRD, capable of generating realistic differentiating databases, and effective at uncovering discrepancies undetected by the official test-based evaluation. 
Future work includes designing richer constraint extraction to support cross-table constraints (potentially leveraging data profiling tools such as Metanome~\cite{Papenbrock:2015:DPM:2824032.2824086}) and devising their SMT encodings, extending the underlying bounded equivalence checker to larger SQL fragments, incorporating user-specified domain knowledge alongside automatic extraction, and investigating constraint-enhanced verification for SQL verification tasks beyond Text-to-SQL evaluation.
 

\newpage
\bibliographystyle{IEEEtran}   
\bibliography{lit}

\newpage
\newpage
\onecolumn
\appendices

\section{\spotitplus}
\label{app:spotitplus}

\subsection{Specification Mining}
Given a database instance $D$ over schema $S$, we extract candidate constraints from the observed data for each of the constraint types.

\paragraph{Range Constraints.}
For numeric columns with at least 2 distinct values, we derive bounds using three complementary approaches: 

\begin{enumerate}
\item \textbf{Strict bounds} use observed minimum and maximum values $[\min(c), \max(c)]$, ensuring every current database value satisfies the constraint.
\item \textbf{Loose bounds} apply Tukey's fence method ($k=3$): $[Q_1 - 3 \cdot \text{IQR},\ Q_3 + 3 \cdot \text{IQR}]$, accommodating outliers while excluding unrealistic extremes. For columns where observed minimum is $\geq 0$, we enforce a lower bound of 0. 
\item \textbf{LLM-generated semantic bounds} infer domain-appropriate ranges from column metadata, observed statistics, and sample values (e.g., ages $[0, 120]$, percentages $[0, 100]$, one-sided bounds $[0, \infty)$ for counts).
\end{enumerate}

During validation, each range constraint is reviewed and one of these three bounding approaches is selected. In our experiments, the LLM validator chose between strict and semantic bounds.

\paragraph{Categorical Constraints.}
For columns with 2-30 unique values:
\begin{enumerate}
\item Extract the complete set of observed values $V = \{v_1, \ldots, v_k\}$.
\item Generate a candidate constraint: $\cin(c, V)$.
\end{enumerate}

Columns with cardinality $> 30$ are automatically excluded to prevent over-constraining the counterexample search space.

\paragraph{Null Constraints.}
For columns with zero observed null values, we generate a candidate constraint. The process of filtering primary key constraints occurs during validation.

\paragraph{Functional Dependencies.}
For each pair of columns $(c_a, c_b)$:
\begin{enumerate}
\item Compute $n_a = |\text{distinct}(c_a)|$ and $n_{ab} = |\text{distinct}(c_a, c_b)|$.
\item Verify that grouping by $c_a$ produces groups with exactly one unique $c_b$ value each.
\item Generate a candidate constraint: $c_a \funcdep c_b$.
\end{enumerate}

\paragraph{Ordering Dependencies.}
For each pair of numeric columns $(c_1, c_2)$:
\begin{enumerate}
\item Test whether $c_1[i] \leq c_2[i]$ holds for all non-null rows.
\item Test whether $c_1[i] \geq c_2[i]$ holds for all non-null rows.
\item If either relationship holds, generate a candidate constraint.
\end{enumerate}

\subsection{Implementation}

Our extraction system is implemented in Python 3.11 and consists of approximately 2,300 lines of code organized into five main modules:

\begin{itemize}
\item \texttt{extract\_constraints\_LLM.py} ($\sim$800 lines): Orchestrates the \LLM constrain extraction pipeline.

\item \texttt{extract\_constraints.py} ($\sim$800 lines): Orchestrates the \ruleBased constraint extraction pipeline.

\item \texttt{utils.py} ($\sim$100 lines): Dependency detection and schema mapping utilities.

\item \texttt{llm.py} ($\sim$300 lines): Language model interaction and prompt management using OpenAI's API.

\item \texttt{prompts.py} ($\sim$600 lines): Structured prompts for semantic bound generation and validation tasks.
\end{itemize}

The system uses OpenAI's \textit{gpt-5.1} model with JSON mode enabled for parsable outputs.

\subsection{Example Prompt}
\label{app:prompt}

We provide an example system prompt for validating categorical constraints.

\begin{lstlisting}[language={}, basicstyle=\small\ttfamily, breaklines=true, frame=single, numbers=none, caption={System prompt for categorical constraint validation.}]
You are a database constraint expert analyzing columns for IN 
constraints (categorical/enumerated values).

Your task is to determine if a column should have an IN 
constraint that restricts values to a specific set.

Consider:
- Is this a true categorical/enumerated column where all 
  valid values are known and limited?
  Examples: status codes, types, categories, yes/no flags
- Or is this pseudo-categorical with potentially more values 
  in the real world?
  Examples: names, product IDs, sparse data
- Does the column description indicate it is an enumeration?
- Could new categories appear in future data?

Respond in JSON format:
{
  "should_constrain": 0 or 1,
  "reasoning": "brief explanation"
}
\end{lstlisting}

\section{Additional Counterexamples}
\label{app:examples}

We show additional examples illustrating the effect of considering database constraints in bounded verification. The first two examples are cases where the database constraints ruled out unrealistic counterexamples. The next two examples show the qualitative differences between counterexamples generated with and without database constraints.

\textbf{Example D.1.} Consider the natural language question: \textit{"How much is the average build up play speed of the Heart of Midlothian team?"}. The generated query computes the average using \texttt{AVG(BUILDUPPLAYSPEED)}, which skips \texttt{NULL} values in both the sum and the count. The gold query instead computes \texttt{SUM(BUILDUPPLAYSPEED) / COUNT(ID)}, where \texttt{SUM} skips \texttt{NULL} values but \texttt{COUNT(ID)} counts all rows. \vanilla finds a counterexample where one \\ \texttt{TEAM\_ATTRIBUTES} row has \texttt{BUILDUPPLAYSPEED = NULL} and the other has \\ \texttt{BUILDUPPLAYSPEED = -2}. Neither \ruleBased nor \LLM finds a counterexample because both extract a \texttt{NOT NULL} constraint on \texttt{BUILDUPPLAYSPEED}, which reflects the actual data and demonstrates that the counterexample that \vanilla finds would be trivial in a real-world scenario.

\begin{figure}[H]
\small
\noindent
\begin{minipage}[t]{0.48\textwidth}
\textbf{Generated Query $P$:}
\begin{lstlisting}[language=SQL, basicstyle=\scriptsize\ttfamily, breaklines=true, escapeinside={(*@}{@*)}]
SELECT (*@\colorbox{pink}{AVG(BUILDUPPLAYSPEED)}@*) 
FROM TEAM_ATTRIBUTES 
JOIN TEAM 
  ON TEAM_ATTRIBUTES.TEAM_API_ID 
     = TEAM.TEAM_API_ID 
WHERE TEAM.TEAM_LONG_NAME 
  = 'HEART OF MIDLOTHIAN';
\end{lstlisting}
\end{minipage}%
\hspace{0.02\textwidth}%
\begin{minipage}[t]{0.48\textwidth}
\textbf{Gold Query $Q$:}
\begin{lstlisting}[language=SQL, basicstyle=\scriptsize\ttfamily, breaklines=true, escapeinside={(*@}{@*)}]
SELECT (*@\colorbox{pink}{CAST(SUM(T2.BUILDUPPLAYSPEED)}@*) 
(*@\colorbox{pink}{AS REAL) / COUNT(T2.ID)}@*) 
FROM TEAM AS T1 
INNER JOIN TEAM_ATTRIBUTES AS T2 
  ON T1.TEAM_API_ID = T2.TEAM_API_ID 
WHERE T1.TEAM_LONG_NAME 
  = 'HEART OF MIDLOTHIAN';
\end{lstlisting}
\end{minipage}
\vspace{0.3cm}

\noindent
\begin{minipage}[t]{0.98\textwidth}
\begin{lstlisting}[language=SQL, basicstyle=\scriptsize\ttfamily, breaklines=true]
--SpotIt Counterexample--
CREATE TABLE TEAM (
    ID INTEGER, TEAM_API_ID INTEGER, TEAM_FIFA_API_ID INTEGER,
    TEAM_LONG_NAME VARCHAR(20), TEAM_SHORT_NAME VARCHAR(20)
);
INSERT INTO TEAM VALUES (-2147483648, -2147483648, -2147483647, 'HEART OF MIDLOTHIAN', '2147483648');
INSERT INTO TEAM VALUES (-2147483647, -2147483648, -2147483648, 'HEART OF MIDLOTHIAN', '2147483648');
CREATE TABLE TEAM_ATTRIBUTES (
    ID INTEGER, TEAM_FIFA_API_ID INTEGER, TEAM_API_ID INTEGER,
    DATE VARCHAR(20), BUILDUPPLAYSPEED INTEGER, ...
);
INSERT INTO TEAM_ATTRIBUTES VALUES (-2147483648, -2147483647, -2147483648, '2147483648', NULL, ...);
INSERT INTO TEAM_ATTRIBUTES VALUES (-2147483647, -2147483647, -2147483648, '2147483648', -2, ...);
\end{lstlisting}
\end{minipage}
\caption{Generated Query $P$, Gold Query $Q$, and \vanilla counterexample with bound $K=2$ for NL Question: \textit{"How much is the average build up play speed of the Heart of Midlothian team?"}.}
\label{fig:cex-1095}
\end{figure}

\newpage
\textbf{Example D.2.} Consider the natural language question: \textit{"What is the most common bond type?"}. The generated query orders by \texttt{COUNT(BOND\_TYPE)}, which skips \texttt{NULL} values, while the gold query orders by \texttt{COUNT(BOND\_ID)}, which counts all rows in each group. \vanilla finds a counterexample where one bond has \texttt{BOND\_TYPE = NULL}. Neither \ruleBased nor \LLM finds a counterexample because both extract a \texttt{NOT NULL} constraint on \texttt{BOND.BOND\_TYPE} from the observed data, where every bond has a recorded type. Under this constrained domain, \texttt{COUNT(BOND\_TYPE)} and \texttt{COUNT(BOND\_ID)} produce identical counts within each group, making the queries equivalent.

\begin{figure}[H]
\small
\noindent
\begin{minipage}[t]{0.48\textwidth}
\textbf{Generated Query $P$:}
\begin{lstlisting}[language=SQL, basicstyle=\scriptsize\ttfamily, breaklines=true, escapeinside={(*@}{@*)}]
SELECT BOND_TYPE 
FROM BOND 
GROUP BY BOND_TYPE 
ORDER BY (*@\colorbox{pink}{COUNT(BOND\_TYPE)}@*) DESC 
LIMIT 1;
\end{lstlisting}
\end{minipage}%
\hspace{0.02\textwidth}%
\begin{minipage}[t]{0.48\textwidth}
\textbf{Gold Query $Q$:}
\begin{lstlisting}[language=SQL, basicstyle=\scriptsize\ttfamily, breaklines=true, escapeinside={(*@}{@*)}]
SELECT T.BOND_TYPE 
FROM ( 
  SELECT 
    BOND_TYPE, 
    COUNT(BOND_ID) 
  FROM BOND 
  GROUP BY BOND_TYPE 
  ORDER BY (*@\colorbox{pink}{COUNT(BOND\_ID)}@*) DESC 
  LIMIT 1 
) AS T;
\end{lstlisting}
\end{minipage}
\vspace{0.3cm}

\noindent
\begin{minipage}[t]{0.98\textwidth}
\begin{lstlisting}[language=SQL, basicstyle=\scriptsize\ttfamily, breaklines=true]
-- SpotIt Counterexample--
CREATE TABLE BOND (
    BOND_ID VARCHAR(20), MOLECULE_ID VARCHAR(20), BOND_TYPE VARCHAR(20)
);
INSERT INTO BOND VALUES ('2147483648', '2147483649', NULL);
INSERT INTO BOND VALUES ('2147483649', '2147483648', '2147483649');
\end{lstlisting}
\end{minipage}
\caption{Generated Query $P$, Gold Query $Q$, and \vanilla counterexample with bound $K=2$ for NL Question: \textit{"What is the most common bond type?"}.}
\label{fig:cex-195}
\end{figure}

\newpage
\textbf{Example D.3.} Consider the natural language question: \textit{"Which citizenship do the vast majority of the drivers hold?"}. The generated and gold queries differ in their \texttt{ORDER BY} clauses, with the generated query using \texttt{COUNT(NATIONALITY)} and the gold query using \texttt{COUNT(DRIVERID)}. Since \texttt{COUNT(column)} skips \texttt{NULL} values while \texttt{COUNT(DRIVERID)} counts all rows (as a primary key, \texttt{DRIVERID} is never \texttt{NULL}), a counterexample requires a driver with \texttt{NATIONALITY = NULL}. Both \vanilla and \LLM produce valid counterexamples exploiting this difference. However, \ruleBased extracts a \texttt{NOT NULL} constraint on \texttt{DRIVERS.NATIONALITY} from the observed data, where every driver has a recorded nationality. This constraint prevents \zthree from assigning \texttt{NULL} to any \texttt{NATIONALITY} value, making the two \texttt{COUNT} expressions equivalent under the constrained domain.

\begin{figure}[H]
\small
\noindent
\begin{minipage}[t]{0.47\textwidth}
\textbf{Generated Query $P$:}
\begin{lstlisting}[language=SQL, basicstyle=\scriptsize\ttfamily, breaklines=true, escapeinside={(*@}{@*)}]
SELECT NATIONALITY 
FROM DRIVERS 
GROUP BY NATIONALITY 
ORDER BY (*@\colorbox{pink}{COUNT(NATIONALITY)}@*) DESC
LIMIT 1;
\end{lstlisting}
\end{minipage}%
\hspace{0.02\textwidth}%
\begin{minipage}[t]{0.47\textwidth}
\textbf{Gold Query $Q$:}
\begin{lstlisting}[language=SQL, basicstyle=\scriptsize\ttfamily, breaklines=true, escapeinside={(*@}{@*)}]
SELECT NATIONALITY 
FROM DRIVERS 
GROUP BY NATIONALITY 
ORDER BY (*@\colorbox{pink}{COUNT(DRIVERID)}@*) DESC 
LIMIT 1;
\end{lstlisting}
\end{minipage}
\vspace{0.1cm}

\noindent
\begin{minipage}[t]{0.47\textwidth}
\begin{lstlisting}[language=SQL, basicstyle=\scriptsize\ttfamily, breaklines=true, escapeinside={(*@}{@*)}]
--SpotIt Counterexample--
CREATE TABLE DRIVERS (
    DRIVERID INTEGER,
    DRIVERREF VARCHAR(20),
    NUMBER INTEGER,
    CODE VARCHAR(20),
    FORENAME VARCHAR(20),
    SURNAME VARCHAR(20),
    DOB DATE,
    NATIONALITY VARCHAR(20),
    URL VARCHAR(20)
);
INSERT INTO DRIVERS VALUES 
  (1, '2147483648', 0, '2147483648',
   '2147483648', '2147483648',
   '0001-01-01', NULL, '2147483648');
INSERT INTO DRIVERS VALUES 
  (0, '2147483648', 0, '2147483648',
   '2147483648', '2147483648',
   '0001-01-01', '2147483649',
   '2147483648');
\end{lstlisting}
\end{minipage}%
\hspace{0.05\textwidth}%
\begin{minipage}[t]{0.47\textwidth}
\begin{lstlisting}[language=SQL, basicstyle=\scriptsize\ttfamily, breaklines=true, escapeinside={(*@}{@*)}]
--SpotIt+ Counterexample--
CREATE TABLE DRIVERS (
    DRIVERID INTEGER,
    DRIVERREF VARCHAR(20),
    NUMBER INTEGER,
    CODE VARCHAR(20),
    FORENAME VARCHAR(20),
    SURNAME VARCHAR(20),
    DOB DATE,
    NATIONALITY VARCHAR(20),
    URL VARCHAR(20)
);
INSERT INTO DRIVERS VALUES 
  (1, '2147483648', 0, '2147483648',
   '2147483648', '2147483648',
   '0001-01-01', NULL, '2147483648');
INSERT INTO DRIVERS VALUES 
  (2, '2147483649', 0, '2147483648',
   '2147483648', '2147483649',
   '0001-01-02', '2147483649',
   '2147483648');
\end{lstlisting}
\end{minipage}
\label{fig:cex-997}
\end{figure}

\newpage
\textbf{Example D.4.} Consider the natural language question: \textit{"List the last name of members with a major in environmental engineering and include its department and college name."}. The generated query is missing the \texttt{POSITION = 'MEMBER'} filter present in the gold query, so it returns all members majoring in Environmental Engineering regardless of their position. All three approaches find valid counterexamples containing members with positions other than \texttt{'MEMBER'} (such as \texttt{'Inactive'} and \texttt{'Secretary'}), which are returned by the generated query but filtered out by the gold query. The realism of the counterexamples varies significantly: \vanilla uses placeholder values like \texttt{'2147483648'} and \texttt{'8982140592376775526'} for positions, while \ruleBased and \LLM produce realistic categorical values such as \texttt{'Secretary'}, \texttt{'Medium'} for t-shirt sizes, and \texttt{'College of Natural Resources'} for the college name. 

\begin{figure}[ht]
\small
\noindent
\begin{minipage}[t]{0.48\textwidth}
\textbf{Generated Query $P$:}
\begin{lstlisting}[language=SQL, basicstyle=\scriptsize\ttfamily, breaklines=true]
SELECT T1.LAST_NAME, T2.DEPARTMENT, 
  T2.COLLEGE 
FROM MEMBER AS T1 
INNER JOIN MAJOR AS T2 
  ON T1.LINK_TO_MAJOR = T2.MAJOR_ID 
WHERE T2.MAJOR_NAME 
  = 'ENVIRONMENTAL ENGINEERING';
\end{lstlisting}
\end{minipage}%
\hspace{0.02\textwidth}%
\begin{minipage}[t]{0.48\textwidth}
\textbf{Gold Query $Q$:}
\begin{lstlisting}[language=SQL, basicstyle=\scriptsize\ttfamily, breaklines=true]
SELECT T2.LAST_NAME, T1.DEPARTMENT, 
  T1.COLLEGE 
FROM MAJOR AS T1 
INNER JOIN MEMBER AS T2 
  ON T1.MAJOR_ID = T2.LINK_TO_MAJOR 
WHERE T2.POSITION = 'MEMBER' 
  AND T1.MAJOR_NAME 
  = 'ENVIRONMENTAL ENGINEERING';
\end{lstlisting}
\end{minipage}
\vspace{0.001cm}

\noindent
\begin{minipage}[t]{0.98\textwidth}
\begin{lstlisting}[language=SQL, basicstyle=\scriptsize\ttfamily, breaklines=true]
--SpotIt Counterexample--
CREATE TABLE MAJOR (MAJOR_ID VARCHAR(20), MAJOR_NAME VARCHAR(20), DEPARTMENT VARCHAR(20), COLLEGE VARCHAR(20));
INSERT INTO MAJOR VALUES ('2147483649', 'ENVIRONMENTAL ENGINEERING', '2147483648', '2147483648');
INSERT INTO MAJOR VALUES (...);
CREATE TABLE MEMBER (MEMBER_ID VARCHAR(20), FIRST_NAME VARCHAR(20), LAST_NAME VARCHAR(20), EMAIL VARCHAR(20),
    POSITION VARCHAR(20), T_SHIRT_SIZE VARCHAR(20), PHONE VARCHAR(20), ZIP INTEGER, LINK_TO_MAJOR VARCHAR(20));
INSERT INTO MEMBER VALUES ('2147483648', '2147483648', '2147483648', '2147483648', '2147483648', '2147483648', '2147483648', 1, '2147483648');
INSERT INTO MEMBER VALUES ('2147483649', '2147483648', '2147483648', '2147483648', '8982140592376775526', '2147483648', '2147483648', 1, '2147483649');
\end{lstlisting}
\end{minipage}
\vspace{0.01cm}
\end{figure}

\begin{figure}[ht]
\noindent
\begin{minipage}[t]{0.98\textwidth}
\begin{lstlisting}[language=SQL, basicstyle=\scriptsize\ttfamily, breaklines=true]
--SpotIt+-NoV Counterexample--
CREATE TABLE MAJOR (MAJOR_ID VARCHAR(20), MAJOR_NAME VARCHAR(20), DEPARTMENT VARCHAR(20), COLLEGE VARCHAR(20));
INSERT INTO MAJOR VALUES ('recdIBgeU38UbV2sy', 'ENVIRONMENTAL ENGINEERING', '2147483649', 'College of Natural Resources');
INSERT INTO MAJOR VALUES (...);
CREATE TABLE MEMBER (MEMBER_ID VARCHAR(20), FIRST_NAME VARCHAR(20), LAST_NAME VARCHAR(20), EMAIL VARCHAR(20),
    POSITION VARCHAR(20), T_SHIRT_SIZE VARCHAR(20), PHONE VARCHAR(20), ZIP INTEGER, LINK_TO_MAJOR VARCHAR(20));
INSERT INTO MEMBER VALUES ('recf4UKTfipCzgcSA', '2147483648', '2147483649', '2147483649', 'Secretary', 'Large', '2147483648', 1020, 'recxRBSgVYeSEGvyo');
INSERT INTO MEMBER VALUES ('recro8T1MPMwRadVH', '2147483649', '2147483648', '2147483648', 'Inactive', 'Small', '2147483649', 1021, 'recdIBgeU38UbV2sy');
\end{lstlisting}
\end{minipage}
\vspace{0.01cm}

\noindent
\begin{minipage}[t]{0.98\textwidth}
\begin{lstlisting}[language=SQL, basicstyle=\scriptsize\ttfamily, breaklines=true]
--SpotIt+ Counterexample--
CREATE TABLE MAJOR (MAJOR_ID VARCHAR(20), MAJOR_NAME VARCHAR(20), DEPARTMENT VARCHAR(20), COLLEGE VARCHAR(20));
INSERT INTO MAJOR VALUES ('2147483649', 'ENVIRONMENTAL ENGINEERING', '2147483649', '2147483649');
INSERT INTO MAJOR VALUES (...);
CREATE TABLE MEMBER (MEMBER_ID VARCHAR(20), FIRST_NAME VARCHAR(20), LAST_NAME VARCHAR(20), EMAIL VARCHAR(20),
    POSITION VARCHAR(20), T_SHIRT_SIZE VARCHAR(20), PHONE VARCHAR(20), ZIP INTEGER, LINK_TO_MAJOR VARCHAR(20));
INSERT INTO MEMBER VALUES ('2147483648', '2147483648', '2147483648', '2147483648', 'Inactive', 'X-Large', '2147483648', 0, '2147483649');
INSERT INTO MEMBER VALUES ('2147483649', '2147483648', '2147483648', '2147483648', 'Secretary', 'Medium', '2147483648', 1, '2147483649');
\end{lstlisting}
\end{minipage}
\caption{Generated Query $P$, Gold Query $Q$, \vanilla, \ruleBased, and \LLM counterexamples with bound $K=2$ for NL Question: \textit{"List the last name of members with a major in environmental engineering and include its department and college name."}}
\label{fig:cex-1426}
\end{figure}

\end{document}